\documentclass[twocolumn,secnumarabic,amssymb, nobibnotes, aps, prd]{revtex4}
\usepackage{amsmath}
\usepackage{graphicx}
\usepackage{dcolumn}
\usepackage{bm}
\usepackage{amssymb}
\usepackage{color}
\usepackage{xcolor}
\usepackage{xfrac}

\newcommand{\be}{\begin{equation}}
\newcommand{\ee}{\end{equation}}

\begin{document}

\title{Comment on ``Universal Lindblad equation for open quantum systems"}
\author{Jae Sung Lee$^{1}$}\email{jslee@kias.re.kr}
\author{Joonhyun Yeo$^{1,2}$} \email{jhyeo@konkuk.ac.kr}
\affiliation{{$^1$School of Physics and Quantum Universe Center, Korea Institute for Advanced Study, Seoul 02455, Korea}\\{$^2$Department of Physics, Konkuk University, Seoul 05029, Korea}}

\date{\today}

\begin{abstract}
{In this Comment, we show that the thermal Gibbs state given in terms of a time-independent system Hamiltonian is not a steady state solution of the quantum master equation introduced by Nathan and Rudner [Phys.\ Rev.\ B \textbf{102}, 115109 (2020)], in contrast to their claim.}
\end{abstract}
\pacs{05}

 \maketitle


Recently, Nathan and Rudner~\cite{main} suggested a new quantum master equation (QME) to overcome the shortcomings of previous QMEs. Their equation is in Lindblad form and, thus, satisfies the completely positive trace-preserving condition, which is not guaranteed in the Bloch-Redfield formalism. Their QME is based on the Born-Markov approximation and is valid when the correlation time of the bath is much shorter than the characteristic time scale of system-bath interactions. As the rotating wave approximation is not required, which is usually used to obtain the conventional Lindblad master equation \cite{Breuer}, they claimed that their equation can be applied to broader physical systems compared with previously proposed QMEs. In addition, the new QME can tackle the problem of time-dependent Hamiltonian systems, which is not straightforward in usual Lindblad equations~\cite{Kosloff}.

With these advantages, the QME developed in Ref.~\cite{main} looks promising for understanding open quantum systems in a universal way as they claimed. In this Comment, we point out that the thermal Gibbs state given in terms of a time-independent system Hamiltonian is \textit{not} a steady state solution of this QME for a system in contact with an equilibrium environment. This is in contrast to the claim made in Ref.~\cite{main}, where a numerical simulation seems to show that the average of an observable coincides in the long time limit with the average obtained from the thermal Gibbs state. We note that the stationary thermal Gibbs state is one of the simplifying features of the conventional Lindblad equation \cite{Breuer}. Below, we show analytically that the Gibbs state is not a stationary solution of the QME. We also present an explanation as to how the numerical simulation was interpreted mistakenly as evidence for the stationary thermal Gibbs state.


For simplicity, we consider a system with a time-independent system Hamiltonian $H_s$ and a single noise channel, i.e.,
$H_\textrm{int}  = \sqrt{\gamma} XB$ following the notation of Ref.~\cite{main}. From Eqs.~(30)-(34) in Ref.~\cite{main},
the QME in the Schr\"{o}dinger picture can be written as
\begin{align}
\partial_t \rho = - i [H_s +\Lambda, \rho ] + L \rho L^\dagger - \frac{1}{2} \{ L^\dagger L, \rho  \}, \label{eq:QME1}
\end{align}
where $L$ and $\Lambda$ are given by
\begin{align}
L &= 2 \pi \sqrt{\gamma} \sum_{m,n} g(E_n-E_m) X_{mn} |m\rangle \langle n|
\label{eq:L}
\end{align}
and
\begin{align}
\Lambda &= \sum_{l,m,n} f (E_l-E_m,E_n-E_l)X_{ml} X_{ln} |m \rangle \langle n| ,
\label{eq:Lambda}
\end{align}
with $f(E_1,E_2) \equiv  - 2 \pi \gamma \mathcal{P}\int_{-\infty}^\infty d \omega ~\omega^{-1} g(\omega-E_1)g(\omega+E_2)$.
Here, $\mathcal{P}$ denotes the Cauchy principal value, $|m\rangle$ is the eigenstate of $H_s$ with eigenvalue $E_m$
and $g$ is a function given in terms of the spectral function of the bath \cite{main,footnote}.

Now, we check whether the thermal state $ \rho^\textrm{th} \equiv e^{-\beta H_s}/\textrm{tr} \{ e^{-\beta H_s} \}$
is a stationary state solution of Eq.~\eqref{eq:QME1}. Using Eqs.~(\ref{eq:L}) and (\ref{eq:Lambda}), it is in fact straightforward to
verify that the right hand side of Eq.~(\ref{eq:QME1}) does not vanish when $\rho^\textrm{th}$ is inserted in place of $\rho$.
We find it, however, more instructive
to present the result in terms of the notations used in the derivation of the conventional Lindblad equation~\cite{Breuer}. We first define the operator
\begin{align}
A(\omega) \equiv \sum_{E_n-E_m = \omega} \Pi_m X \Pi_n,
\end{align}
where $\Pi_m = |m \rangle \langle m|$ is the projection operator. Then, $L$ can be rewritten as
\begin{align}
L  &= 2 \pi \sqrt{\gamma} \sum_{m,n} g(E_n-E_m) \Pi_m X \Pi_n \nonumber \\
&=  2 \pi \sqrt{\gamma} \sum_{\omega} \sum_{E_n-E_m =\omega} g(E_n-E_m) \Pi_m X \Pi_n \nonumber \\
&= 2\pi \sqrt{\gamma} \sum_{\omega} g(\omega) A(\omega)
\end{align}
Using the relations $A^\dagger (\omega) = A(-\omega)$ and $g^* (\omega) = g(\omega)$, $L^\dagger$ is given by
\begin{align}
L^\dagger = 2 \pi \sqrt{\gamma} \sum_{\omega} g(\omega) A^\dagger (\omega).
\end{align}
We can also rewrite $\Lambda$ as
\begin{align}
\Lambda &= \sum_{l,m,n,h} f(E_l-E_m,E_n-E_h) \Pi_m X \Pi_l \Pi_h X \Pi_n \nonumber \\
&= \sum_{\omega_1,\omega_2} f(\omega_1,\omega_2) A(\omega_1) A(\omega_2) . \label{eq:Gamma2}
\end{align}

By using the relations satisfied by the thermal state, $\rho^\textrm{th} A(\omega) = e^{\beta \omega} A(\omega) \rho^\textrm{th}$, $\rho^\textrm{th} A^\dagger(\omega) = e^{-\beta \omega} A^\dagger (\omega) \rho^\textrm{th}$, we can show that
\begin{align}
L \rho^\textrm{th} L^\dagger  - & \frac{1}{2} \{ L^\dagger L, \rho^\textrm{th} \}
=-2\pi^2 \gamma \sum_{\omega_1, \omega_2}  \left(1-e^{(\omega_2-\omega_1)/2} \right)^2  \nonumber \\
&~~~~~~~~~~~~~ \times g(\omega_1)g(\omega_2) A^\dagger(\omega_1) A(\omega_2) \rho^\textrm{th}. \label{eq:result}
\end{align}
and
\begin{align}
[\Lambda, \rho^\textrm{th}]  &=  \sum_{\omega_1,\omega_2} f(\omega_1,\omega_2) [ A(\omega_1) A(\omega_2),  \rho^\textrm{th}]  \label{eq:result1}\\
&= \sum_{\omega_1,\omega_2} f(\omega_1,\omega_2) (1-e^{\beta(\omega_1+\omega_2)})A(\omega_1) A(\omega_2)  \rho^\textrm{th}, \nonumber
\end{align}
where we used the Kubo-Martin-Schwinger (KMS) condition, $g(-\omega) = e^{-\beta/2} g(\omega)$ satisfied by the bath.
It is clear that the above two quantities do not vanish in general. It is interesting to note that the above expressions look very similar to those that
appear in the proof of the stationary thermal state for the conventional Lindblad equation \cite{Breuer}. In fact, the expressions for the conventional case
essentially correspond to taking only the diagonal terms in the double sums
($\omega_1=\omega_2$ in Eq.~(\ref{eq:result}) and $\omega_1=-\omega_2$ in Eq.~(\ref{eq:result1})).
The presence of the off-diagonal terms in the double sums
prevents $\rho_\mathrm{th}$ from being a steady state solution.


To confirm our conclusion, we re-examine the simulation done in Section V A of Ref.~\cite{main} for the relaxation process of the magnetization of a spin-chain system. We use the same parameter set as used in Ref.~\cite{main}: $B_z = 8 \eta$, $\Lambda = 100 \eta$, $\omega_0 = 2\eta$, $T_1 = 2 \eta$, and $\gamma_1 = 0.1 \eta$. Here, we set $\gamma_2 = 0$, so that the spin chain is disconnected from the reservoir with temperature $T_2$. The system is initially in the state with all spins aligned against $B_z$. As carried out in Ref.~\cite{main}, we ignore the Lamb shift term $\Lambda$. With these conditions, we obtain numerically the density matrix by directly solving the QME for the spin chain with length $N = 6$. We note that the simulation result is essentially the same, regardless of the length of the spin chain.

\begin{figure}
\resizebox{0.90\columnwidth}{!}{\includegraphics{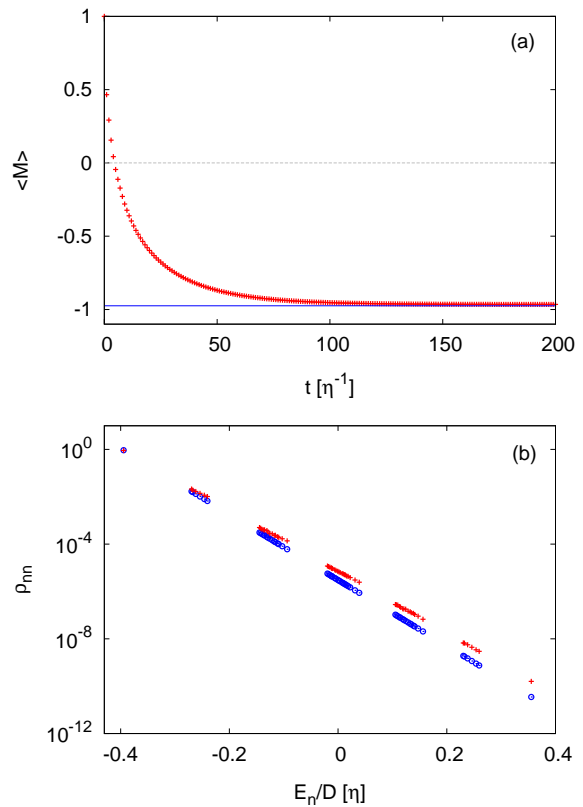}}
\caption{ (a) Plot for $\langle M \rangle$ as a function of time. Blue solid line denotes $\langle M \rangle_\textrm{ss}^\textrm{th}$, the expectation value of $M$ in the Gibbs state at temperature $T_1$. The data points are $\langle M \rangle$ obtained by solving the QME directly. (b) Diagonal density-matrix elements $\rho_\textrm{nn}$ as a function of energy eigenvalues of the system Hamiltonian. $\circ$ and $+$ denote $\rho_\textrm{nn}^\textrm{th}$ and $\rho_\textrm{nn}$, respectively.   }
\label{fig1}
\end{figure}

We first calculate the expectation value of $z$-magnetization $M = \frac{1}{N} \sum_{n = 1}^{N} S_n^z$ in the relaxation process. Figure~\ref{fig1}(a) shows the plot for the expectation value $\langle M \rangle = \textrm{tr}[\rho(t)M]$ as a function of time. We can see that its steady state value $\langle M \rangle_\textrm{ss}$ is very close to $\langle M \rangle_\textrm{ss}^\textrm{th}$, which is the expectation value of $M$ in the Gibbs state at temperature $T_1$. This is in agreement with the result in Ref.~\cite{main}. However, we find that, although the two average values seem to coincide with each other, the steady state itself is, in fact, different from the Gibbs state given, in terms of the system Hamiltonian. Figure~\ref{fig1}(b) shows the plot for the diagonal elements of the steady-state density matrix $\rho_\textrm{nn}=\langle n| \rho|n\rangle$ as a function of the energy eigenvalues of the system Hamiltonian. As the figure shows, the steady state of the QME clearly deviates from the Gibbs state $\rho_\textrm{nn}^\textrm{th} $. Note that $\rho_\textrm{11}$ corresponding to the lowest energy eigenvalue is very close to $\rho_\textrm{11}^\textrm{th}$. The closeness of the two averages, $\langle M \rangle_\textrm{ss} \approx \langle M \rangle_\textrm{ss}^\textrm{th}$, as shown in Fig.~\ref{fig1}(a) and in Ref.~\cite{main}, can be understood from the fact that $\rho_\textrm{11}$ is a dominating term in the numerical evaluation of averages.

From the similar simulations with other parameter sets (not shown here), we find that $\rho_\textrm{nn}$ tends to approach $\rho_\textrm{nn}^\textrm{th}$ for higher temperatures and lower values of $\gamma$. This may imply that the conditions for this QME to have \emph{practically} the thermal state as a steady state solution would be much stronger than those suggested by the authors. Otherwise, the deviation of the steady state from the thermal state of $H_s$ may be interpreted as a strong-coupling effect between the system and the bath. The QME of Ref.~\cite{main} is obtained through a series of Markovian approximations in the weak-coupling limit. It will be interesting to investigate in more detail how the validity conditions for their approximations allow for the strong-coupling effect to emerge in their QME.

Authors acknowledge the Korea Institute for Advanced Study for providing computing resources 
(KIAS Center for Advanced Computation Linux Cluster System).
This research was supported by NRF Grants No.~2020R1F1A1062833
(J.Y.), and individual KIAS Grants No.~QP064902 (J.S.L.) at the Korea Institute for Advanced Study.


\begin{thebibliography}{}
\bibitem{main} F. Nathan and M. S. Rudner, Universal Lindblad equation for open quantum systems. Phys. Rev. B \textbf{102}, 115109 (2020).

\bibitem{Breuer} H.-P. Breuer and F. Petruccione, \emph{The theory of open quantum systems} (Oxford University Press, Great Clarendon Street, 2002).


\bibitem{Kosloff} R. Dann, A. Levy, and R. Kosloff, Time-dependent Markovian quantum master equation. Phys. Rev. A \textbf{98}, 052129 (2018).


\bibitem{footnote} In the course of writing this Comment, we discovered that the first argument of $f$ in Eq.~(34) of \cite{main}, which gives $\Lambda$,
should read
$E_l-E_m$ instead of $E_m-E_l$ in accordance with Eq.~(D3). Equation (D8) contains similar typos.
The definition of $f(E_1,E_2)$ given just below Eq.~(34) should contain an overall minus sign as in Eq.~(D7).





\end{thebibliography}
{}
\end{document}